\documentclass[a4paper, onecolumn, 11pt, unpublished]{quantumarticle}
\pdfoutput=1
\usepackage[utf8]{inputenc}
\usepackage[english]{babel}
\usepackage[T1]{fontenc}
\usepackage{amsmath}
\usepackage{amsfonts}
\usepackage{booktabs}
\usepackage{hyperref}
\usepackage{graphicx}
\usepackage{braket}
\usepackage[numbers]{natbib}
\usepackage{subcaption}
\usepackage{float}
\usepackage{array}
\usepackage{changepage}
\usepackage{tabularx}
\usepackage{multirow}

\usepackage{tikz}
\usepackage{lipsum}
\begin{document}

\title{A Comparison of Quadratic and Higher-Order Representations for QAOA}

\author{Kristina Bell}
\email{kristinab.1506@gmail.com}
\author{Adam Lowe}
\email{adam.lowe@qinetiq.com}
\author{Emily Coles}
\author{Nathanael Ridgway}
\author{Gillian Marshall}
\affiliation{QinetiQ, Malvern Technology Centre, St Andrew's Road, Malvern, WR14 3PS, UK}

\begin{abstract}
  In this work we consider a routing problem and compare quadratic and higher-order representations using the Quantum Approximate Optimisation Algorithm (QAOA). The majority of works investigating QAOA use quadratic Hamiltonians to represent the considered problems, which can lead to poor scaling in qubit requirements. We address the gap of direct comparisons between quadratic and higher-order forms through an investigation into two distinct formulations of the same use case. We find that the higher-order form yields better solution quality and scales better in terms of numbers of qubits, but requires more two-qubit gates. We additionally consider a factoring method to reduce the gate depth of the higher-order version, which achieves a significant reduction in the number of two-qubit gates when run on real IBM hardware.
\end{abstract}


\section{Introduction}

A key field in which quantum computing is theorised to have advantages over classical computing, is in combinatorial optimisation. This class of problems is often NP-hard or even NP-complete, making it intractable on classical computers for large problem sizes. Furthermore, many industry problems fall into this category, including routing \cite{harwood2021formulating, cattelan2024modeling, papalitsas2019qubo}, scheduling \cite{schworm2023solving}, and packing type problems \cite{cellini2024qal}, giving it broad applicability.

A popular way of representing combinatorial optimisation problems for quantum computers is as a Quadratic Unconstrained Binary Optimisation (QUBO) model \cite{glover2019quantum}, which is equivalent to the quadratic Ising model \cite{lucas2014ising}. QUBOs are compatible with both quantum annealing and the gate-based quantum computing paradigm, making them highly versatile \cite{denkena2021quantum, plewa2021variational}. The QUBO model can be extended to a Higher-Order Unconstrained Binary Optimisation (HUBO) model\footnote{HUBOs are sometimes called HOBOs or PUBOs in the literature.} \cite{farhi2014quantum}, which, while being fully compatible with the standard gate-based algorithms, has been studied much less than QUBOs.

One category of combinatorial optimisation problems that is particularly interesting is routing problems. There are a variety of different types of routing problem that have been considered as QUBOs, such as the Vehicle Routing Problem with Time Windows (VRPTW) \cite{harwood2021formulating}, the ride pooling problem (RPP) \cite{cattelan2024modeling}, and the Travelling Salesperson Problem (TSP) \cite{papalitsas2019qubo}. They present an interesting application since not only do we need to consider the graph structure of the problem, but additionally there is often a timing element that adds an extra dimension to consider. This can lead to very poor scaling in the number of qubits required to encode the problem, motivating the use of HUBOs for a representation with better scaling. There has been some exploration of HUBOs for routing problems, mainly for TSP \cite{salehi2022unconstrained, glos2022space}, which this work builds on.

In this work, we compare a QUBO and HUBO representation for a routing problem, inspired by a real problem, involving traversing a graph collecting assets, with an overall goal of reaching an end point within a set time limit. We call this problem the Asset Retrieval Problem (ARP). We consider the qubit scaling of both formulations, as well as the gate depth and two-qubit gate count of the naive gate implementation, through experimentation on IBM hardware. We run the two approaches on several small example problems to understand the solution quality and thus analyse which approach is more appropriate for this problem, under current hardware constraints.

Within gate-based quantum computing, the most common algorithm for solving QUBOs is the Quantum Approximate Optimisation Algorithm (QAOA) \cite{farhi2014quantum}. QAOA is a hybrid quantum-classical algorithm that is feasible for running on near-term quantum devices and shows promise for demonstrating quantum advantage in the future \cite{boulebnane2024solving}. It works by constructing a quantum ansatz that alternates cost and mixer Hamiltonians to form an approximation of the problem. A classical optimiser is then used to update the parameters of the cost and mixer Hamiltonians to reach the minimum energy state, which is then the problem solution (where problems are minimisations by convention).

There has been much research aimed at understanding QAOA and increasing its utility \cite{blekos2024review}, prompting the development of variants such as Recursive QAOA (R-QAOA) \cite{bravyi2020obstacles} that runs QAOA recursively, fixing a subset of parameters at a time, and Warm-Start QAOA \cite{egger2021warm} that involves solving a relaxed problem and using that solution as the initial starting state of QAOA. Additionally, newer algorithms have started to emerge that adapt the QAOA model to mitigate some of its limitations. For example, the Quantum Alternating Operator Ansatz (QAOA+ for this paper, although often just QAOA) \cite{hadfield2019quantum} uses the same principle of the QAOA ansatz construction, but allows for more customisable mixers. This means, for example, that a constraint can be encoded directly into the mixer, theoretically restricting the search space to the feasible subspace of that constraint.

Unlike quantum annealing \cite{kim2025quantum}, QAOA, and its variants, are not limited to quadratic-only problems and are able to support higher-order terms.  Despite this, there has been limited research directly comparing QUBO and HUBO inputs to QAOA. There has been some work done on constructing HUBOs for specific problems \cite{ salehi2022unconstrained, glos2022space, sano2024accelerating, campbell2022qaoa, fuchs2021efficient, domino2022quadratic}, but this is often done in isolation, or with restrictions. For purely HUBO exploration, there have been investigations aimed at understanding how QAOA \cite{sachdeva2024quantum} and QAOA+ \cite{pelofske2024scaling} behave, particularly in relation to quantum annealing \cite{sachdeva2024quantum, gilbert2023solving}, but a direct comparison with QUBOs is often not attempted. In cases where both a QUBO and HUBO is constructed, this is often done through either converting the QUBO to a HUBO via converting one-hot encoding of variables to binary encoding \cite{sano2024accelerating, tabi2020space}, or by mapping the HUBO to a QUBO via quadratization \cite{domino2022quadratic, pelofske2024short, grange2024quadratic, stein2023evidence}. This leads to situations where one encoding has better scaling, both in terms of qubit count and circuit depth, and where the other resulting encoding may not be in its most efficient form. Consequently, there is still limited understanding of how QUBO and HUBO formulations compare when neither is a more obvious choice.

Since a HUBO can support higher-order terms, it has the potential to be able to represent problems more compactly, as can be seen in \cite{salehi2022unconstrained, campbell2022qaoa, domino2022quadratic}. However, in general, this can come at the cost of requiring more two-qubit gates, since higher-order terms involve more inter-connected qubits, and thus multi-controlled gates that need to be decomposed into two-qubit gates to fit onto current quantum computers. This restriction to at most two-qubit gates may change in the future, with research suggesting ways of implementing higher order gates in various paradigms \cite{lu2019global, glaser2023controlled, dlaska2022quantum}. However, in the current Near-Term Intermediate Scale (NISQ) era, minimising the number of two-qubit gates remains highly important, since they are a significant contributor of noise to the system, and can thus be an influential source of errors. Thus, the trade-off between qubit scaling and two-qubit gate scaling needs to be explored and understood to inform on the feasibility of the implementation.

Additionally, although the HUBO incurs many two-qubit gates due to the higher-order terms, it is possible to reduce the number by using techniques proposed in \cite{campbell2022qaoa} and \cite{sivarajah2020t} to perform cancellations. One of our key results is demonstrating the effectiveness of these techniques in significantly improving the two-qubit gate count of our circuits.

Overall, we find that although the HUBO scales worse than the QUBO in terms of gate depth and two-qubit gate count, the improvement in qubit scaling is sufficient to yield a higher success probability and increased solution quality on IBM hardware. This highlights that the HUBO formulation may be a more practical formulation of certain problems on gate-based quantum computers.

This paper is organised as follows. In Section 2 we provide a brief overview of QAOA and detail the factoring method that was used to reduce the number of two-qubit gates in the HUBO implementation. In Section 3 we introduce the routing problem that was investigated, along with the QUBO and HUBO formulations for the problem. In Section 4 we summarise the experimentation methodology and the test cases. In Section 5 we provide the results of the experimentation. We conclude with a discussion of the results and direction for future work in Section 6.


\section{QAOA Overview}
QAOA \cite{farhi2014quantum} is a hybrid quantum-classical algorithm, in which quantum circuits are generated with parameters that are  optimised by a classical optimiser. In theory, by partitioning up the problem into quantum and classical processes, the depth of the quantum circuits can be smaller than would otherwise be needed, making them less prone to errors and thus more likely to be feasible on a NISQ-era device. 

QAOA can be considered as the gate-based analogue of quantum annealing. The quantum circuit is initialised in the ground state of a known system: a uniform superposition generated using Hadamard gates. The system is then slowly evolved towards the ground state of the target problem by applying a suitably generated ansatz, then measuring to sample from the probability distribution of the resulting state.

In QAOA, this ansatz is constructed using alternating parameterised cost and mixer operators. We start with a cost Hamiltonian, $\hat{H}_C$, that encodes the problem to be solved, and is different for every problem. For QUBOs where at most quadratic terms are present, a problem with $n$ variables can be represented as a symmetric $n\times n$ matrix $Q$, so that the cost Hamiltonian is defined as
\begin{align}
\hat{H}_C := \frac{1}{4}\sum_{i,j=1}^n Q_{ij}Z_iZ_j - \frac{1}{2}\sum_{i=1}^n \big(\sum_{j=1}^nQ_{ij}\big)Z_i,
\end{align}
for $Z_i$ denoting the application of a Pauli-$Z$ gate to qubit $i$.
A mixer Hamiltonian, $\hat{H}_M$, is conventionally defined as
\begin{align}
\hat{H}_M := \sum_{i=1}^n X_i,
\end{align}
where $X_i$ denotes the application of a Pauli-$X$ gate to qubit $i$.

The parameterised cost and mixer operators are then given by the unitary matrices generated through the mappings $\hat{H}_C \rightarrow U_C(\gamma)=e^{-i\gamma\hat{H}_C}$ and $\hat{H}_M \rightarrow U_M(\beta)=e^{-i\beta\hat{H}_M}$, where the Pauli gates get exponentiated into rotation gates: $e^{-i\theta Z} = R_Z(2\theta)$ and similarly for $X$. The overall QAOA ansatz is given by alternating these operators for $p$ layers, with $p \geq 1$, so that the overall circuit is:
\begin{align}
    U = U_M(\beta_p)U_C(\gamma_p)...U_M(\beta_1)U_C(\gamma_1)H^{\otimes n}\ket{0}
\end{align}
with $2p$ parameters $\{\gamma_1,...\gamma_p\}, \{\beta_1,...\beta_p\}$, and $H$ being the Hadamard gate.

Theoretically, the quality of the approximation should increase as we use more layers, however, this leads to deeper circuits, so practical QAOA is generally currently limited to around 1-2 layers.

As is standard in quantum computing, each time a quantum circuit needs to be run, it is evaluated multiple times due to the inherent stochastic nature of quantum algorithms. This generates a probability distribution that is the outcome of that run of the quantum circuit. The cost value of the distribution is calculated and used to update the parameters for the next cycle of the optimisation loop. The parameters are controlled using a classical optimiser, that eventually halts after the optimiser converges or a set number of cycles is completed. The mode of the final probability distribution is taken as the solution. For additional detail see \cite{farhi2014quantum}.

A QUBO over variables $x_i \in \{0,1\}$ has the general form:
\begin{align}
\text{minimise} \sum_{i}a_{i}x_{i} + \sum_{i\neq j}b_{i,j}x_{i}x_{j}
\end{align}
for $a_{i}$ and $b_{i,j}$ coefficients specific to the problem.

A HUBO is defined similarly to a QUBO, but allowing for higher-order terms. In general, to translate a problem from QUBO/HUBO form into an ansatz for QAOA, the binary variables are mapped to Ising spin variables $z_i \in \{-1,1\}$ by the bijective map $z_i = 2x_i - 1$. The spin variables are then equivalent to Pauli $Z$ Hamiltonians, which are exponentiated as in the QUBO case to give $R_{Z}(2\theta)$ operators. The non-linear $z_1...z_k$ entries map to multi-controlled $R_{Z_1...Zk}(2\theta)$ operators, consisting of sequences of $CNOT$ gates pre- and post-pending an $R_Z(2\theta)$ operator.

\begin{figure}[h!]
    \begin{subfigure}{0.55\linewidth}
        \centering
        \includegraphics[width=\linewidth]{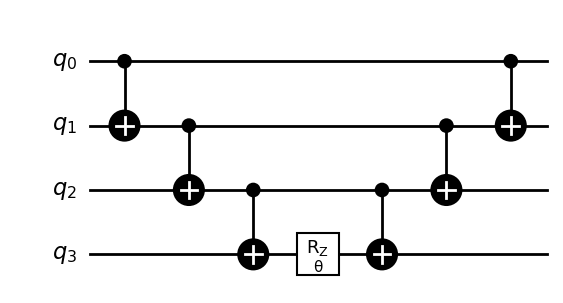}
        \caption{A ladder phase gadget.}
        \label{fig:ladder_pg}
    \end{subfigure}
    \hfill
    \begin{subfigure}{0.43\linewidth}
        \centering
        \includegraphics[width=\linewidth]{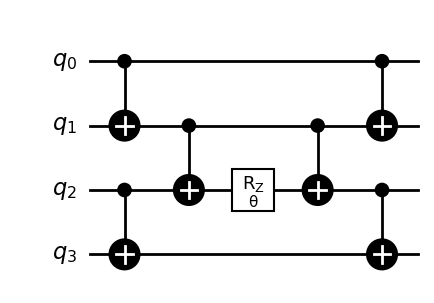}
        \caption{A tree phase gadget.}
        \label{fig:tree_pg}
    \end{subfigure}
    \caption{Examples of the most common ways of constructing phase gadgets.}
    \label{fig:phase_gadgets}
\end{figure}

These multi-controlled $RZ$ operators are known as Phase Gadgets \cite{sivarajah2020t}, and can be constructed in a variety of ways, as can be seen in Figure \ref{fig:phase_gadgets}. The only requirement is that the $CNOT$ gates span all of the relevant qubits, and that the $RZ$ gate is on the target of the central $CNOT$. This means that the number of $CNOT$ gates required for a term with $k$ variables is $2(k-1)$, which can lead to high two-qubit gate counts for high-ordered terms.

However, since $CNOT$ gates are self-inverse, it is possible to construct these phase gadgets such that, when appropriately ordered, cancellations can be made, thus reducing the total number of two-qubit gates in the circuit. Identifying the optimal ordering is an NP task, but a factoring heuristic can be used to create an efficient ordering that, while possibly not optimal, is highly effective. This heuristic is mentioned in \cite{campbell2022qaoa}, and originally taken from \cite{sivarajah2020t}.

\begin{figure}[h!]
    \begin{subfigure}{0.58\linewidth}
        \centering
        \includegraphics[width=\linewidth]{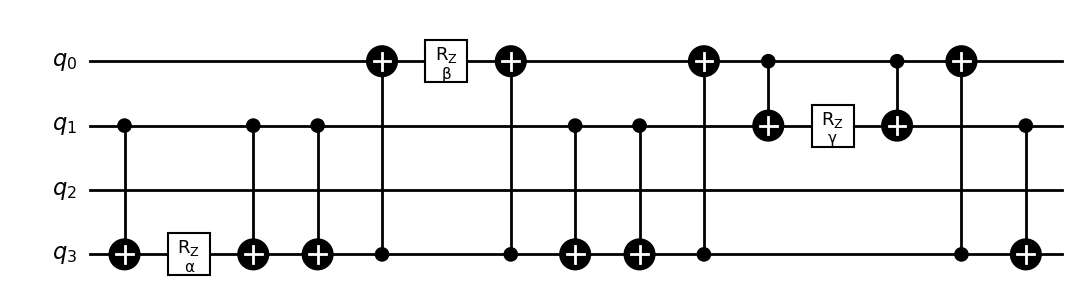}
        \caption{A phase gadget construction for the example set before \\ factoring.}
        \label{fig:example_nofactor}
    \end{subfigure}
    \hfill
    \begin{subfigure}{0.38\linewidth}
        \centering
        \includegraphics[width=\linewidth]{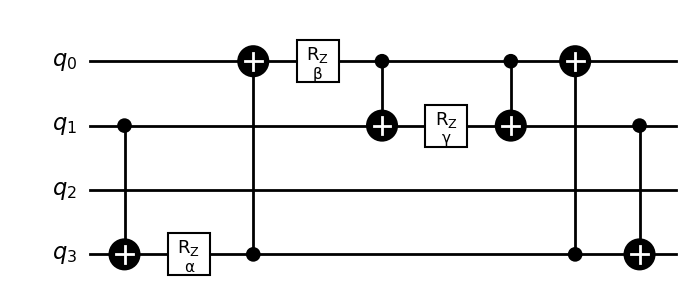}
        \caption{A phase gadget construction for the example set after factoring.}
        \label{fig:example_factor}
    \end{subfigure}
    \hfill
    \caption{An example of factoring the phase gadget construction for the set $\{x_1x_2x_3x_4, x_1x_2x_4, x_2x_4\}$. $\alpha$, $\beta$, and $\gamma$ denote rotation parameters that would be defined at algorithm runtime.}
    \label{fig:phase_gadget_example}
\end{figure}

It involves greedily grouping the terms in the HUBO by shared variables, such that each successively shorter term is a subset of the variables in the longer terms. For example, given a collection of terms $\{x_1x_2x_3x_4, x_1x_2x_3, x_1x_2x_4, x_2x_4, x_1x_4\}$, we begin by selecting the longest term $\{x_1x_2x_3x_4\}$. We then select one of the next longest terms that is a subset of our initial term, giving the updated set $\{x_1x_2x_3x_4, x_1x_2x_4\}$. Finally, we choose a next longest term that is a subset of all of our terms (which is equivalent to being a subset of the most recently added term). This gives the grouping $S=\{x_1x_2x_3x_4, x_1x_2x_4, x_2x_4\}$. By definition of the grouping, the $CNOT$ construction for each term is a subset of the $CNOT$ construction for the next longest term, meaning that the entire group requires only as many $CNOT$ gates as the overall longest term in the group. An example for the set $S$ is shown in Figure \ref{fig:phase_gadget_example}.

\section{Problem Implementation}

We consider a scenario involving an agent tasked with collecting assets. The agent is initially located at $A$, and has the requirement of reaching location $B$ within the required time limit $T \in \mathbb{N}$. Along its possible routes are asset locations from which it can collect assets, under the condition that it still reaches $B$ within the time limit.

The problem is thus: What is the optimal route the agent can take such that it maximises the number of assets collected en route to $B$?

We can visualise this scenario as a network graph, with the nodes numbered $1, ..., n$, denoting the other asset locations. We will refer to these nodes as \textbf{internal nodes}, denoted as $I_n$. The overall set of nodes in the graph is then $N=\{A\}\cup I_n \cup \{B\} = \{A, 1, ..., n, B\}$. The value of each edge denotes the time it takes to traverse that edge, and the value of each node denotes the number of assets at that node. Note that the time is restricted to integers to allow for practical QUBO and HUBO representation.

\begin{figure}[H]
    \centering
    \includegraphics[width=0.9\linewidth]{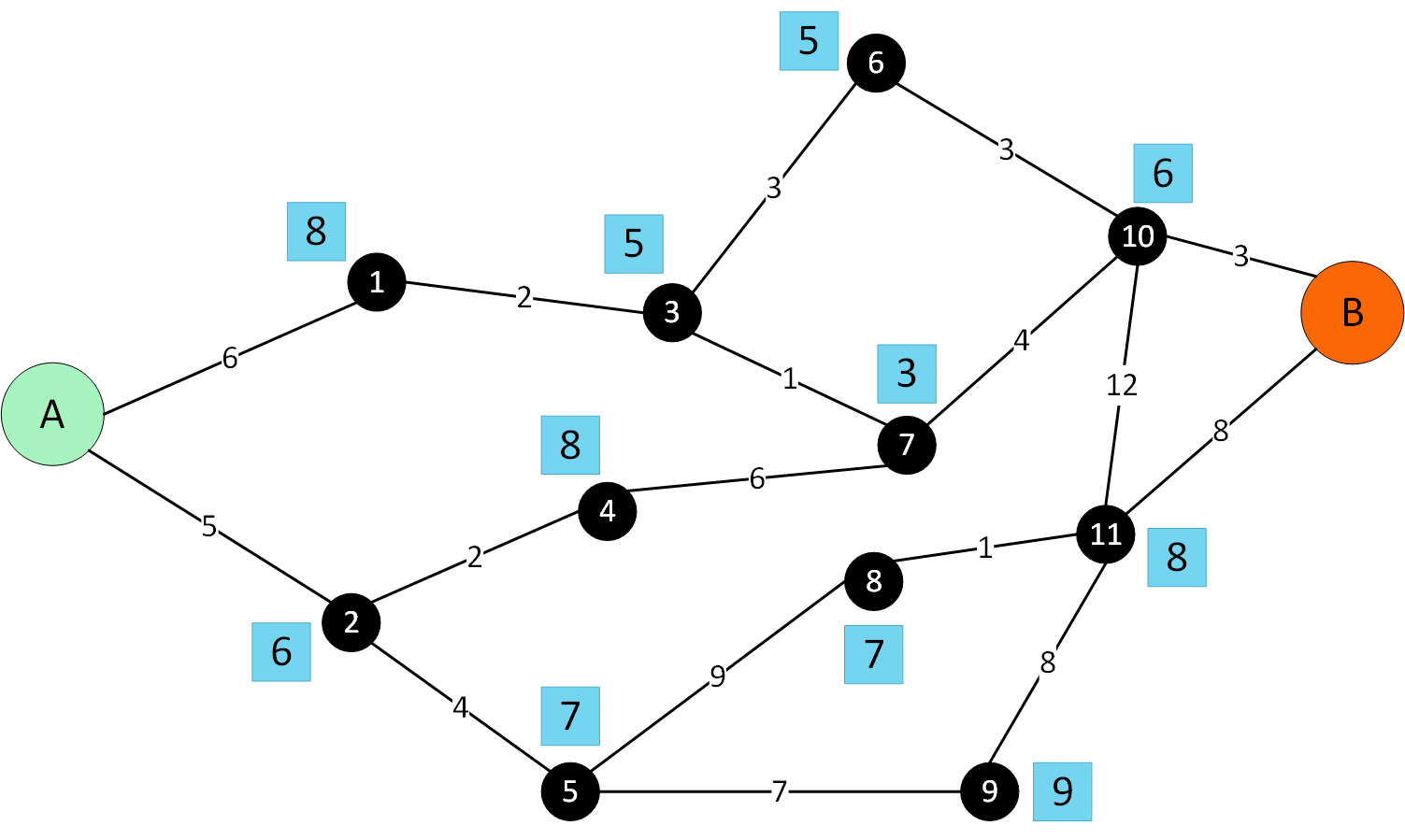}
    \caption{An example of the asset routing problem for a large number of nodes. This example contains 11 internal nodes and 17 edges. We can choose $T=20$ to allow sufficient time to traverse the graph, whilst providing multiple feasible routes.}
    \label{fig:example_scenario}
\end{figure}

An example scenario is shown in Figure \ref{fig:example_scenario}, where we choose $T=20$ to allow enough time to traverse the graph whilst providing several route options for doing so. The numbers on the edges represent the time it takes to traverse that edge. Additionally, each internal node will have some value attached to it, which is the value of the assets to be retrieved at that node.

We now step through the two formulations for the general use case. We first provide the QUBO formulation, then provide the HUBO formulation. In both cases, the overall equation includes a constraint penalty parameter, $\alpha$. This parameter is used to ensure that violations of the constraint do not improve the objective enough to be seen as worthwhile. In simple cases, it can be calculated analytically, while in more complex cases, such as this one, it is often found through experimentation.

\subsection{QUBO Formulation}

We first define the route that the agent takes as a path $p_0, ..., p_S$, for $S\leq T$ total steps. By allowing the agent to remain at $B$ for multiple steps at the end of its route, we can require that the agent performs exactly $T$ steps, fixing $S=T$ and simplifying the path calculations without loss of generality.

\bigskip

We define the binary variables $x_{u,v}^{(i)} \in \{0,1\}$ of the QUBO in terms of edges per time-step as follows:
\begin{align}
x_{u,v}^{(i)} = 
    \begin{cases} 
    1 & \mbox{if } p_{i-1}=u, p_{i}=v \\ 
    0 & \mbox{otherwise}  
    \end{cases}
\end{align}
where $u,v \in N$ and $i \in \{1, ..., T\}$.

The objective is to maximise the number of collected assets, i.e. the values of the visited internal nodes. We define $c_v$ to be the asset value of node $v$, i.e. the 'cost' of node $v$.

In order to be able to accurately calculate if a node has been visited, we require that each internal node may be visited at most once, giving the objective as
\begin{align}\label{objective}
    \mbox{maximise  } \sum_{i=1}^{T}\sum_{u\in N}\sum_{v\in I_n} c_v x_{u,v}^{(i)}.
\end{align}

\noindent Note that this requires that the 'internal graph' (the graph of internal nodes only) be fully connected so that retracing of steps is implicitly allowed. This can be ensured by using Dijkstra's shortest path algorithm over all nodes to generate the shortest paths to each node. Since Dijkstra's runs at $\mathcal{O}(|N|log|N|)$, the total time would be $\mathcal{O}(|N^2|log|N|)$, which does not overwhelm the overall algorithm.

By convention, QUBOs are minimisation problems so the objective is multiplied by $-1$ to generate the objective within the QUBO construction:
\begin{align}\label{qubo_}
    \mbox{minimise  } Q_0 = -\sum_{i=1}^{T}\sum_{u\in N}\sum_{v\in I_n} c_v x_{u,v}^{(i)}.
\end{align}
There are four main constraints that need to be included within the formulation:
\begin{enumerate}
\item $Q_{1}$: Each internal node can be visited at most once; 
\item $Q_{2}$: The agent must traverse exactly one edge at a time;
\item $Q_{3}$: The path must be completed within the time limit;
\item $Q_{4}$: Each edge must follow from the previous.
\end{enumerate}

$Q_1$ is calculated by summing over the internal nodes and requiring that for each pair of variables arriving at the same internal node, only one of them can be non-zero:

\begin{align}\label{qubo_Q1}
    Q_1 = \sum_{v\in I_n}\sum_{i,j}\sum_{\substack{a,b \\ (a,i) \neq (b,j)}} x_{a,v}^{(i)} x_{b,v}^{(j)}
\end{align}
where $i,j$ are over all possible steps, $a,b$ are over all possible start nodes, i.e. $a,b \in I_n \cup \{A\}$.

For $Q_2$, we can reduce the number of variables required in the formulation by identifying that the path must start at $A$ and end at $B$, so we can separate the sum into three parts. Then, the requirement is given by the sum over edges being exactly one for each time-step:

\begin{equation}\label{qubo_Q2}
\begin{aligned}
    Q_2 = & \sum_{i=2}^{T-1}\Big( \sum_{u,v} x_{u,v}^{(i)}-1\Big)^2 + \Big( \sum_v x_{A,v}^{(1)}-1\Big)^2 + \Big( \sum_u x_{u,B}^{(T)}-1\Big)^2 \\
    \equiv & \sum_{i=2}^{T-1} \Big(\sum_{u,v}\sum_{\substack{(a,b) \\ \neq (u,v)}}x_{u,v}^{(i)}x_{a,b}^{(i)} - \sum_{u,v}x_{u,v}^{(i)} \Big) \\ & + \sum_v\sum_{b \neq v} x_{A,v}^{(1)} x_{A,b}^{(1)} - \sum_v x_{A,v}^{(1)} + \sum_u \sum_{a \neq u} x_{u,B}^{(T)} x_{a,B}^{(T)} -\sum_u x_{u,B}^{(T)}.
\end{aligned}
\end{equation}
We use the $\equiv$ notation to indicate that we are dropping the constant from the equation. Since this is an optimisation problem, the optimal assignment is independent of any constant, so constants are dropped to simplify the computations. The QUBO and HUBO have different constructions, with different constant terms. This means that, despite the fact that they are representing the same problem instances, the optimal energy value for the QUBO and HUBO differs due to the dropping of the constants.

$Q_3$ is an inequality, so must be converted into an equality through the use of slack variables. A common way of defining slack variables is by using a binary encoding \cite{glover2019quantum}. Since we are keeping our use cases small, we can use the simpler one-hot encoding without leading to a significant increase in the number of variables. Thus, we define variables $z_j \in \{0,1\}$, for $j \in \{0, ..., T\}$, that describe the difference between the path length and $T$, i.e.
\begin{equation*}
    z_j = \begin{cases}
        1 & \mbox{if } |\mbox{path}| + j = T \\
        0 & \mbox{otherwise.}
    \end{cases}
\end{equation*}
These have an additional constraint that only one can be non-zero:
\begin{equation}
    \begin{aligned}
        Q_5 = & \Big( \sum_j z_j - 1 \Big)^2 \\
        \equiv & -\sum_j z_j + \sum_j \sum_{k\neq j} z_jz_k,
    \end{aligned}
\end{equation}
With these additional variables, we can write the $Q_3$ constraint as:
\begin{equation}\label{qubo_Q3}
    \begin{aligned}
        Q_3 = & \Big(\sum_{i,u,v}t_{u,v}x_{u,v}^{(i)} + \sum_{j}jz_j - T\Big)^2 \\
        \equiv & \sum_{i,u,v} \big(t_{u,v}^2 - 2Tt_{u,v}\big) x_{u,v}^{(i)} + \sum_j \big(j^2-2Tj\big) z_j \\ & + \sum_{i,u,v}\sum_{\substack{(i',a,b)\\ \neq (i,u,v)}}t_{u,v}t_{a,b}x_{u,v}^{(i)}x_{a,b}^{(i')} + \sum_{j}\sum_{k\neq j}jkz_jz_k + \sum_{i,u,v}\sum_j 2jt_{u,v}x_{u,v}^{(i)}z_j.
    \end{aligned}
\end{equation}
The final constraint, $Q_4$, is given by:
\begin{equation}\label{qubo_Q4}
    \begin{aligned}
        Q_4 = & \sum_{i,v} \Big(\sum_a x_{a,v}^{(i-1)} - \sum_b x_{v,b}^{(i)}\Big)^2 \\
        = & \sum_{i,v} \bigg[\sum_a x_{a,v}^{(i-1)} + \sum_a\sum_{u\neq a} x_{a,v}^{(i-1)}x_{u,v}^{(i-1)} + \sum_b x_{v,b}^{(i)} + \sum_b\sum_{u \neq b} x_{v,b}^{(i)} x_{v,u}^{(i)} - 2\sum_a\sum_b x_{a,v}^{(i-1)}x_{v,b}^{(i)}\bigg].
    \end{aligned}
\end{equation}
This gives the overall QUBO of
\begin{equation*}
    Q = Q_0 + \alpha Q_1 + \alpha Q_2 + \alpha Q_3 + \alpha Q_4 + \alpha Q_5,
\end{equation*}
for a suitable constraint penalty $\alpha$. $\alpha$ was found empirically by testing a range of values on small examples in simulation. These values were chosen by noting that they needed to be roughly around the maximal possible value of any node, $\max(c_v)$. Consequently, we took values that were a proportion of $\max(c_v)$. From this, $\alpha$ was set to $0.75 \times \max(c_v)$, (with $c_v$ the node values) as this was found to yield optimal solutions with the highest probability.

\subsection{HUBO Formulation}

The HUBO is constructed similarly to the QUBO, but with a different variable definition that allows for more favourable scaling of number of qubits required. The path and cost of nodes are defined as they were for the QUBO.
For the HUBO, we define binary variables $x_u^{(i)}$ as:
\begin{equation}
x_u^{(i)} = 
    \begin{cases} 
    1 & \mbox{if } p_{i}=u \\ 
    0 & \mbox{otherwise}  
    \end{cases}
\end{equation}
where $u \in N\symbol{92}\{A\}$ and $i \in \{1, ..., T-1\}$. Note that we can ignore $u=A$ because we know that $A$ is only touched at time $T=0$. Similarly, time $i=T$ isn't required, since time $T$ can only be touched at node $B$.

The objective is equivalent to the QUBO case:
\begin{equation}\label{hubo_Q0}
    H_0 = -\sum_{i=1}^{T-1}\sum_{u\in I_n} c_u x_u^{(i)}.
\end{equation}

The HUBO also has four main constraints. The first three constraints are the same as the QUBO case, while the fourth is unique to the HUBO. Again, we introduce slack variables with their additional constraint to construct $Q_3$.

\begin{enumerate}
\item $H_{1}$: Each internal node can be visited at most once; 
\item $H_{2}$: The agent must traverse exactly one edge at a time;
\item $H_{3}$: The path must be completed within the time limit;
\item $H_{4}$: Each node must be connected by an edge to the next (note that this can be restricted to only apply to $B$ by limiting the $\text{time-step}=1$ variables to neighbours of $A$, since all internal nodes are connected) and once the agent reaches $B$ it cannot leave.
\end{enumerate}

This gives the following mathematical expressions for the constraints (where $x_A^0$ and $x_B^T$ are $1$ implicitly and for any other $u$ we have $x_u^0 = 0$ and $x_u^T = 0$). For $H_4$ we have used $N(B)$ to denote the neighbourhood of node $B$, i.e. any nodes that share an edge with $B$. This is as follows:

\begin{flalign}\label{hubo_Q1}
    H_1 = \sum_{u\in I_n}\sum_{i\neq j} x_u^{(i)} x_u^{(j)};&&
\end{flalign}

\begin{flalign}
&\begin{aligned}
    H_2 = & \sum_i \Big( \sum_u x_u^{(i)} - 1 \Big )^2 \\
    \equiv & \sum_i \Big( \sum_u \sum_{v \neq u} x_u^{(i)} x_v^{(i)} - \sum_u x_u^{(i)} \Big );
\end{aligned}&&
\end{flalign}

\begin{flalign}
    \begin{aligned}
        H_3 = & \Big( \sum_{i=1}^{T}\sum_{\substack{u,v \\ u \neq v}} t_{u,v} x_u^{(i-1)}x_v^{(i)} \Big)^2 + \Big( \sum_j jz_j \Big )^2 + T^2 \\ 
        &+ 2 \Big (\sum_{i=1}^{T}\sum_{\substack{u,v \\ u \neq v}} t_{u,v} x_u^{(i-1)}x_v^{(i)} \Big)\Big( \sum_j jz_j \Big ) - 2T\sum_{i=1}^{T}\sum_{\substack{u,v \\ u \neq v}} t_{u,v} x_u^{(i-1)}x_v^{(i)} - 2T\sum_j jz_j \\
        \equiv & \sum_{i=1}^{T}\sum_{\substack{u,v \\ u \neq v}} (t_{u,v}^2 - 2Tt_{u,v}) x_u^{(i-1)}x_v^{(i)} + \sum_j (j^2-2Tj)z_j + \sum_j\sum_{k \neq j} jkz_jz_k \\
        &+ 2 \sum_{i=1}^{T}\sum_j\sum_{\substack{u,v \\ u \neq v}} jt_{u,v} x_u^{(i-1)}x_v^{(i)} z_j + \sum_{\substack{i,u,v \\ u \neq v}}\sum_{\substack{i',u',v' \\ u' \neq v' \\ (i,u,v) \neq (i',u',v')}} t_{u,v}t_{u',v'} x_u^{(i-1)}x_v^{(i)}x_{u'}^{(i'-1)}x_{v'}^{(i')};
    \end{aligned}&&
\end{flalign}

\begin{flalign}\label{hubo_Q4}
    H_4 = \sum_{i > 1} \sum_{u \notin N(B)} x^{(i-1)}_u x^{(i)}_B +
    \sum_{i < T-1} \sum_{u \neq B}  x^{(i)}_B x^{(i+1)}_u;&&
\end{flalign}

\begin{flalign}
    \begin{aligned}
        H_5 = & \Big( \sum_j z_j - 1 \Big)^2 \\
        \equiv & -\sum_j z_j + \sum_j \sum_{k \neq j} z_jz_k.
    \end{aligned}&&
\end{flalign}

This gives the overall HUBO of
\begin{equation*}
    H = H_0 + \alpha H_1 + \alpha H_2 + \alpha H_3 + \alpha H_4 + \alpha H_5,
\end{equation*}
for a suitable constraint penalty $\alpha$. $\alpha$ was set to $0.75 \times \max(c_v)$, (with $c_v$ the node values) as this was found to yield the best results, when small cases were tested in simulation using the same approach as for the QUBO case.


\section{Experimental Methodology}
From the above problem formulations, it was straightforward to generate the quantum circuits required for QAOA using the techniques from Section 2. However, the scaling of the number of qubits required for the problem means that even trivial scenarios quickly become infeasible for running on existing quantum computers or in simulation. Consequently, some practical steps needed to be taken to allow for experimentation.

Firstly, some efficient classical pre-processing methods were performed to reduce the number of qubits to something that could be simulated or run on real hardware. In particular, using some simple graph algorithms, such as Dijkstra's or Minimum Spanning Tree, several of the QUBO or HUBO variables could be determined to be infeasible, either because the position could not be reached by that time, or because that combination of position and time would make it impossible to finish the route within the time limit. They could thus be removed from the overall objective. This was sufficient to enable us to run some small test cases on real hardware.

Additionally, as discussed in Section 2, a heuristic factoring method was implemented to test if the two-qubit gate count could be meaningfully reduced. In our experimentation, we considered not just the HUBO and QUBO approaches, but additionally the HUBO approach with factoring. Since the HUBO approach tends to yield a significantly higher gate depth than the QUBO approach due to the high two-qubit gate count required to construct the higher-order terms, it is unlikely that a HUBO method will be used without some circuit optimisation, so testing a simple optimisation method would produce a more realistic comparison of the approaches.

Quantum computers are currently limited both by the number of qubits and by gate depth capacity. Consequently, the example provided in Figure \ref{fig:example_scenario} is infeasible to be run on the currently available quantum devices. Instead, in order to test the code and verify the correctness of the QUBO and HUBO formulations from section 3, four test cases were considered. The first test case used two internal nodes for 4 timesteps, the second used three internal nodes for 5 timesteps, the third used three internal nodes for 6 timesteps, and the fourth used four internal nodes for 6 timesteps. Diagrams of the test cases can be found in Appendix \ref{app:test_cases}.

To conduct a fair experiment, we fixed the hyper-parameters of the QAOA experimentation. The number of layers, $p$ was fixed at 1 to minimise the impact due to noise. The classical optimiser was chosen to be the Cobyla optimiser as this was found to perform best in preliminary testing. To reduce the influence of randomness on the circuit measurement, each run of QAOA was performed for 1024 shots. Finally, each test case was repeated ten times to provide averages for each metric, as well as explore the individual performances.

An additional optimisation of QAOA was to calculate the final solution using a different approach than the standard one of taking the mode of the final probability distribution. At each stage of the optimisation loop, a probability distribution is generated and used to inform the next stage of the cycle. This involves calculating the cost value of each of the measured assignments. For this variation of QAOA, the assignment with the minimal cost value at each stage is stored. Then, once the optimisation loop is complete, we return the assignment with the overall minimal cost over all stages\footnote{This approach is inspired by the Qiskit 0.40 QAOA implementation's `best measurement' feature.}. This is guaranteed to be at least as good as the mode by definition, and adds no extra effort since the cost values have to be calculated regardless.

For the experimentation, all code was written using Python and Jupyter Notebooks. Experimentation on real hardware was performed on the 156 qubit IBM Fez device.

\section{Results}

To compare the different approaches, aside from the result quality, there were three key factors that needed to be considered. Firstly, as has been already addressed, the number of qubits required should be minimised. Secondly, the total gate depth of the circuit should be minimised as this is directly correlated with the noise, and thus error. Thirdly, the number of two-qubit gates should be minimised, as two-qubit gates are often the most significant contributor of error to a qubit circuit. For our experimentation, the gate depth and number of two-qubit gates was calculated after the circuit had been transpiled for the specific device.

QAOA is a highly stochastic process, with the variance being directly correlated with the size of the problem, meaning that the confidence in the declared solution accuracy decreases with problem size. Consequently, multiple ways of calculating solution quality were used. Firstly, the average difference between the found result and the classically calculated optimal solution, normalised over the optimal solution, for the ten repeats of each test, to account for stochasticity. Secondly, for each test, the ten repeats were considered separately to identify any outliers and to see how many times an optimal solution was found.

Full tables of results can be found in Appendix \ref{app:tables}, in Table \ref{tab:qubo_results}, Table \ref{tab:hubo_results}, and Table \ref{tab:hubo_f_results}, for the QUBO, HUBO, and factored HUBO (HUBO\_F) respectively. Results on hardware runs are summarised in the next sections.

\subsection{Scaling Behaviour}

The qubit requirements are shown in Table \ref{tab:q_count}. Note that since the factoring impacts gates only, the factored and unfactored HUBO forms had the same qubit requirement. In terms of theoretical scaling, the number of QUBO variables scale according to $O(N^2T)$ while the HUBO variables scale according to $O(NT)$. This is improved by the techniques we employed, but we can still see that the HUBO form required significantly fewer qubits than the QUBO form, and scaled preferably. This likely extends to more general problems, since a HUBO would only be preferable when it allows for a more compact representation.

\begin{table}
    \centering
    \begin{tabular}{ccc}
        \textbf{Test Number} & \textbf{QUBO} & \textbf{HUBO} \\
        \midrule
        1 & 18 & 11\\
        2 & 28 & 13\\
        2 & 38 & 18\\
        4 & 53 & 20\\
    \end{tabular}
    \caption{Qubit counts for each test case}
    \label{tab:q_count}
\end{table}

\begin{figure}[H]
\centering
\includegraphics[width=10.5 cm]{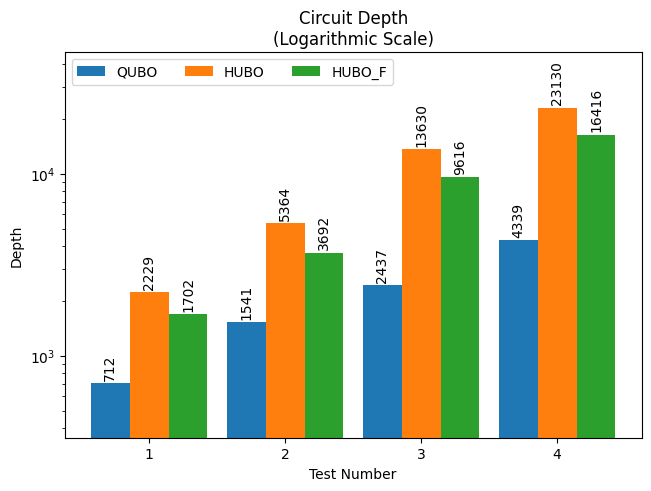}
\caption{Average circuit depth for each test case, averaged across the ten runs for each test case. Shown on a logarithmic scale due to large range. The QUBO has significantly shorter circuits than the HUBO, although the factoring does improve the HUBO circuits.}
\label{circuit_depth}
\end{figure}

Figure \ref{circuit_depth} shows the depth of the circuits averaged over the test cases. From this it can be clearly seen that the QUBO generated favourable circuits, likely due to the high parallelisability of their 2-qubit gates, since higher-order gates are much more likely to have overlap in the qubits used. The factoring did provide some improvement in the circuit depth, but this was still much higher compared to the QUBO depth. In the worst case of the largest test case, the HUBO generated circuits that had roughly 6x the depth of the QUBO circuits, while the factoring improved this to around 4x.

\begin{figure}[H]
\centering
\includegraphics[width=10.5 cm]{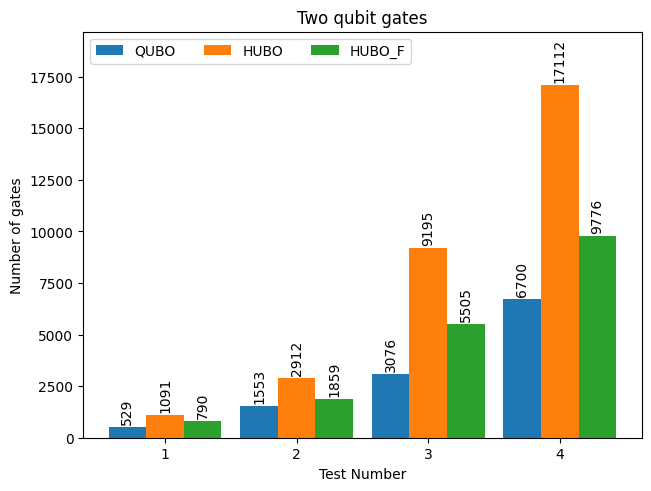}
\caption{Average number of two qubit gates for each test case, averaged across the ten runs for each test case. The factoring yields a significant improvement in the number of two-qubit gates for the HUBO formulation, but does not have better scaling than the QUBO.}
\label{two_qubit_count}
\end{figure}

The two-qubit gates, however, became much closer when the factoring is performed, as shown in Figure \ref{two_qubit_count}. As the problem size increased, the HUBO scaling was still worse than the QUBO scaling for two-qubit gate count, but the factoring provided a significant improvement. For the largest test case, the number of two-qubit gates in the HUBO circuit was roughly 3x the QUBO circuit, but the factoring almost halved this. A more sophisticated factoring method may be able to further improve this such that, for large problems, the fully factored HUBO has comparable two-qubit gates to the QUBO.

The relationship between number of two-qubit gates and circuit depth between the QUBO and HUBO implementations supports the theory that the QUBO is much more parallelisable than the HUBO. For the HUBO, the number of two-qubit gates is always lower than the overall circuit depth. On the other hand, for the QUBO, it is more dependent on the specific test case. The number of two-qubit gates in the QUBO is higher than the overall circuit depth for the two larger cases, slightly lower for the smallest case, and roughly equal for Test Case 2. We can infer from this behaviour that as the problem size increases, the number of two-qubit gates will generally be higher than the circuit depth for the QUBO. It could be argued that this is due to the number of qubits, with more qubits allowing for more parallelisation. However, we note that even for Test Case 1 where the QUBO has 18 qubits, and Test Case 3 where the HUBO has 18 qubits, the difference between circuit depth and two-qubit gates is greater for the HUBO. Consequently, the number of qubits cannot be the only factor, and the quadratic nature of the QUBO must allow for even more space efficient circuit constructions. This suggests that, for cases where the number of gates in QUBO and HUBO constructions are compared analytically, that may not be an accurate reflection of circuit depth relationships.

\subsection{Comparison of Solution Quality}

For the solution quality, we calculated the difference between the solution found on the real hardware, $x_i$ for $i\in[1,10]$, and the optimal solution, $x_{\text{opt}}$, normalised over the optimal solution, averaged over the ten repeats. The equation is given by the normalised distance:
\begin{align}
   N_D = \frac{1}{10} \sum_{i=1}^{10} \frac{|x_i - x_{\text{opt}}|}{|x_{\text{opt}}|}.
\end{align}

\begin{figure}[H]
\centering
\includegraphics[width=10.5 cm]{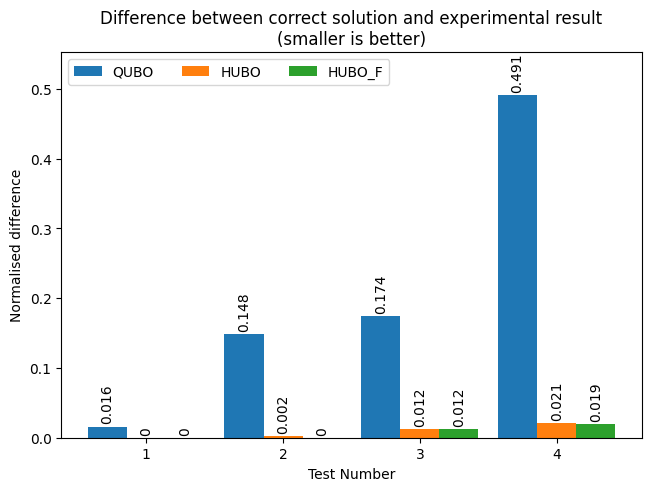}
\caption{Difference between objective function energy of found result and optimal solution, normalised by dividing by correct solution. Averaged over the ten runs of each test case. The QUBO performs much worse than the HUBO (both factored and unfactored), demonstrating that the HUBO is the preferred approach for this application.}
\label{bm_diff}
\end{figure}

We see from the results in Figure \ref{bm_diff} that the HUBO (and factored HUBO) always performed better than the QUBO, with this improvement increasing as the problem size increased. The difference between the HUBO and the factored HUBO was negligible, despite the decrease in circuit depth. We note that the number of qubits for the QUBO in Test Case 1 and the HUBO in Test Case 3 are the same, and the accuracy was comparable. This strongly suggests that the number of qubits is the significant factor in converging on an optimal solution, rather than circuit depth.

\begin{figure}[H]
    \begin{subfigure}{0.45\linewidth}
        \centering
        \includegraphics[width=\linewidth]{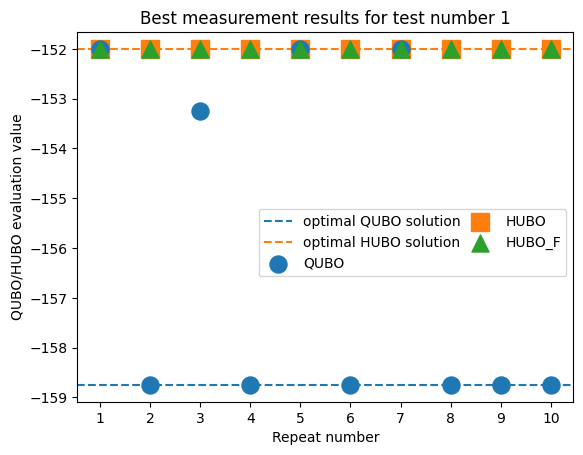}
        \caption{Objective function energies for the found results of test case 1}
        \label{fig:test1_scatter}
    \end{subfigure}
    \hfill
    \begin{subfigure}{0.45\linewidth}
        \centering
        \includegraphics[width=\linewidth]{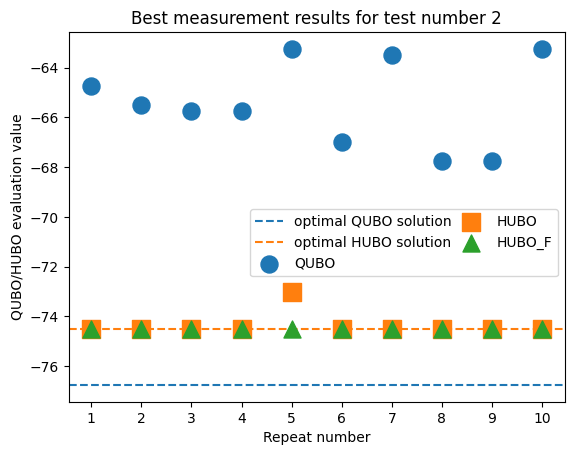}
        \caption{Objective function energies for the found results of test case 2}
        \label{fig:test2_scatter}
    \end{subfigure}
    \hfill
    \begin{subfigure}{0.45\linewidth}
        \centering
        \includegraphics[width=\linewidth]{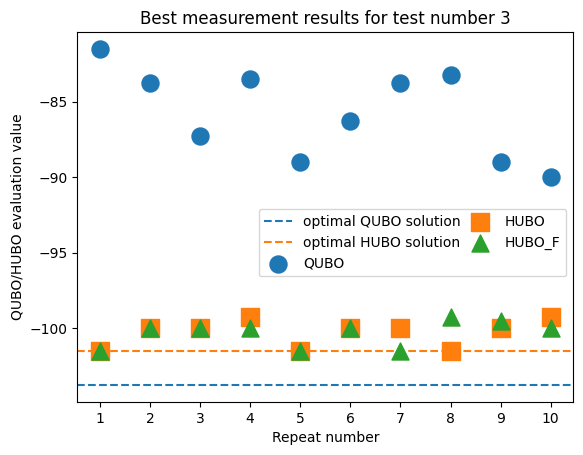}
        \caption{Objective function energies for the found results of test case 3}
        \label{fig:test3_scatter}
    \end{subfigure}
    \hfill
    \begin{subfigure}{0.45\linewidth}
        \centering
        \includegraphics[width=\linewidth]{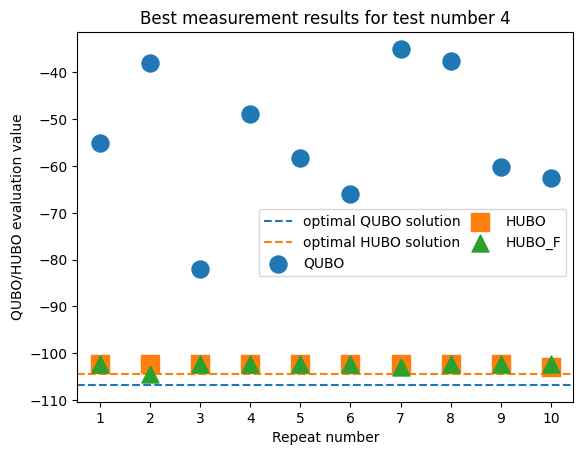}
        \caption{Objective function energies for the found results of test case 4}
        \label{fig:test4_scatter}
    \end{subfigure}
    \caption{Scatter plots showing the objective function energies for the individual repeats of each test case. We can see that the HUBO and factored HUBO perform consistently on or very close to the optimal solution, while the QUBO becomes more scattered, and further from the optimal, by test cases 3 and 4.}
    \label{fig:scatter}
\end{figure}

To further investigate the solution quality, we can compare the individual repeats for each test case. Figure \ref{fig:scatter} shows scatter plots of the ten repeats for each test case separately. A dashed line denotes the optimal solution for the QUBO (blue) and the HUBO and factored HUBO (orange), to show how close each result was. The HUBO and factored HUBO were consistently on or within 2\% of the optimal solution. The QUBO was reasonably consistent for the smallest test case, with only a few outliers, but degraded quickly. The QUBO for Test Case 1 had a standard deviation of only 3.34. By test Case 4, with 53 qubits, the scatter plot for the QUBO appears random, with a standard deviation of 14.8, compared to the HUBO standard deviation of 0.24 and the factored HUBO standard deviation of 0.72.

Figure \ref{fig:scatter} further emphasises that, for this use case at this scale, the HUBO is the clear choice of formulation, due to its improvement in number of qubits. Another benefit of the HUBO formulation is the consistency of the results, which increases confidence in the approach, as well as repeatability of experiments.

\section{Discussion}

This work has shown that there are advantages and disadvantages to both the HUBO and QUBO approaches, but the HUBO outperforms the QUBO on the currently available hardware. The HUBO formulation had greater circuit depths, but smaller circuit widths, and performed significantly better overall. This suggests that, at the current maturity of devices, for this problem, circuit width was a more significant factor than circuit depth. Further experimentation would be needed to analyse how much more compact a HUBO would need to be, in general, to outweigh the increased gate depth requirement, although this is likely to be unique for each problem. The relationship between circuit width and depth is well-established, with many quantum algorithms being able to use shorter circuits if additional ancilla qubits are employed \cite{bennakhi2024analyzing}. One avenue for future work would be a more general assessment of what problem classes benefit most from reduced qubit formulations.

Classical pre-processing has been shown to be a very important tool in making NISQ algorithms more near-term achievable. In this experimentation, simple graph processing steps were taken to reduce the problems to sizes that could be tested, and a classical factoring method was used to simplify the circuits and mitigate against errors caused by noise. There is a growing amount of research into optimisations of algorithms for implementations on NISQ devices, with work being done to bring algorithms to the available devices, rather than wait for the devices to be able to support these algorithms. This includes error mitigation techniques such as Zero Noise Extrapolation \cite{giurgica2020digital}, as well as circuit optimization techniques, such as SWAP strategies \cite{weidenfeller2022scaling}. This work supports these activities, focusing on high-level algorithmic and pre-processing improvements that can be made to reduce the required quantum effort. 

Despite the factoring approach significantly reducing the two-qubit gate count, it is a greedy approach so may not produce an optimal solution. An alternative approach that could be considered in future work uses a parity network \cite{amy2017cnot} instead of a standard quantum circuit representation. These abstract objects may allow for reducing the number of two-qubit gates down to only two per term, regardless of the order of the term, using the approach detailed in \cite{fellner2023parity}. In addition to this, an improved encoding of the variables may yield a more effective HUBO implementation. The use of Gray Code \cite{dominguez2023encoding, di2021improving} for more efficient problem encodings has been shown to be effective in reducing qubit count for Hamiltonian representations, so may be another avenue for improving the HUBO implementation in future iterations.

This work has utilised a realistic use case to investigate two independently derived formulations of the same problem, and compare how the higher-order approach performs against the quadratic approach. We have shown that, for this use case, the HUBO outperforms the QUBO in terms of solution quality, despite having deeper circuits and more two-qubit gates. Consequently, we have found that the number of qubits is the leading contributor to performance, since the HUBO has much fewer qubits than the QUBO. Thus, although the majority of research into quantum approaches to optimisation has focused on quadratic implementations, we conclude that higher-order formulations may provide an avenue for increased optimal performance, and therefore significant research is needed into how they perform in realistic scenarios.

\section*{Acknowledgements}
The authors acknowledge the support of the UK Scientific Technologies Facilities Council (STFC) in providing useful discussion and IBM quantum computing hours for experimentation. In particular, we would like to thank Oscar Wallis and Stefano Mensa. The authors would also like to thank Colin Campbell from Infleqtion for the helpful discussion of his paper `QAOA of the Highest Order' \cite{campbell2022qaoa} and the factoring method used in it.

This work has been part-funded by DSTL under the RCloud framework and through QinetiQ internal investment.

This work was supported by the Hartree National Centre for Digital Innovation, a UK Government-funded collaboration between STFC and IBM. IBM, the IBM logo, and www.ibm.com are trademarks of International Business Machines Corp., registered in many jurisdictions worldwide. Other product and service names might be trademarks of IBM or other companies. The current list of IBM trademarks is available at www.ibm.com/legal/copytrade.

\section*{Author Contributions}
Conceptualization, K.B., E.C., and N.R.; methodology, K.B.; software, K.B., E.C., and N.R.; validation, K.B., E.C., and N.R.; formal analysis, K.B.; investigation, K.B., E.C., and N.R.; writing---original draft preparation, K.B.; writing---review and editing, K.B., A.L., and G.M.; visualization, K.B.; All authors have read and agreed to the published version of the manuscript.

\bibliographystyle{quantum}
\bibliography{bibliography}

\appendix

\section*{Appendices}

\section{Test Cases \label{app:test_cases}}

\subsection{Test Case 1}

\begin{figure}[H]
\centering
\includegraphics[width=0.8\linewidth]{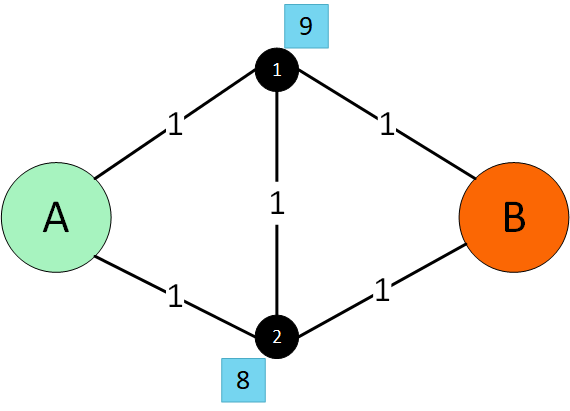}
\caption{Undirected graph for Test Case 1. The values on the edges denote number of time-steps required to traverse that edge. The blue squares denote the asset values at the respective nodes.}
\label{test_case_1}
\end{figure}

Test Case 1 was a graph with two internal nodes, with 4 timesteps. The graph is given in Figure \ref{test_case_1}.

\subsection{Test Cases 2 and 3}

\begin{figure}[H]
\centering
\includegraphics[width=0.8\linewidth]{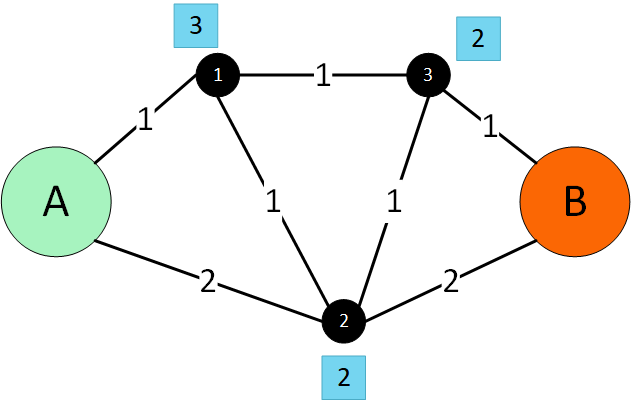}
\caption{Undirected graph for Test Case 2 and Test Case 3. Edge values denote number of timesteps needed to traverse that edge. Blue squares denote node values.}
\label{test_case_2}
\end{figure}

Test Cases 2 and 3 used the same graph with three internal nodes, but with different numbers of timesteps. Test Case 2 used 5 timesteps, while Test Case 3 used 6 timesteps. The graph is given in Figure \ref{test_case_2}.

\subsection{Test Case 4}

\begin{figure}[H]
\centering
\includegraphics[width=0.8\linewidth]{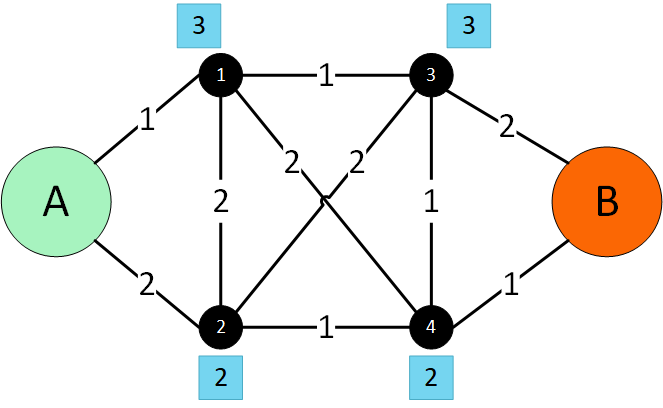}
\caption{Undirected graph for Test Case 4. Edge values denote number of timesteps needed to traverse that edge. Blue squares denote node values.}
\label{test_case_3}
\end{figure}

Test Case 4 used a graph with 4 internal nodes, with 6 timesteps. This graph is given in Figure \ref{test_case_3}.

\onecolumn
\section{Tables of Complete Results \label{app:tables}}

\newcolumntype{C}{>{\centering\arraybackslash}X} 

\begin{table}[H]
\begin{tabularx}{\textwidth}{CCCCC}
\toprule
\textbf{Test No.} & \textbf{Repeat No.}	& \textbf{Circuit Depth} & \textbf{2-qubit gates} & \textbf{Found solution *}\\
\midrule
\multirow[m]{10}{*}{1}
& 1 & 701 & 524 & -152.0\\
& 2 & 685 & 525 & -158.75\\
& 3 & 674 & 521 & -153.25\\
& 4 & 690 & 528 & -158.75\\
& 5 & 697 & 520 & -152.0\\
& 6 & 717 & 537 & -158.75\\
& 7 & 744 & 537 & -152.0\\
& 8 & 764 & 528 & -158.75\\
& 9 & 722 & 530 & -158.75\\
& 10 & 726 & 540 & -158.75\\
\midrule
\multirow[m]{10}{*}{2}
& 1 & 1615 & 1551 & -64.75\\
& 2 & 1580 & 1563 & -65.5\\
& 3 & 1547 & 1556 & -65.75\\
& 4 & 1474 & 1549 & -65.75\\
& 5 & 1653 & 1553 & -63.25\\
& 6 & 1443 & 1545 & -67.0\\
& 7 & 1473 & 1556 & -63.5\\
& 8 & 1585 & 1534 & -67.75\\
& 9 & 1526 & 1557 & -67.75\\
& 10 & 1513 & 1567 & -63.25\\
\midrule
\multirow[m]{10}{*}{3}
& 1 & 2344 & 3062 & -81.5\\
& 2 & 2444 & 3059 & -83.75\\
& 3 & 2475 & 3043 & -87.25\\
& 4 & 2426 & 3082 & -83.5\\
& 5 & 2329 & 3093 & -89.0\\
& 6 & 2505 & 3117 & -86.25\\
& 7 & 2366 & 3028 & -83.75\\
& 8 & 2506 & 3091 & -83.25\\
& 9 & 2418 & 3123 & -89.0\\
& 10 & 2559 & 3060 & -90.0\\
\midrule
\multirow[m]{10}{*}{4}
& 1 & 4070 & 6669 & -55.0\\
& 2 & 4599 & 6741 & -38.0\\
& 3 & 4259 & 6659 & -82.0\\
& 4 & 4512 & 6702 & -49.0\\
& 5 & 4505 & 6688 & -58.25\\
& 6 & 4339 & 6750 & -66.0\\
& 7 & 4265 & 6690 & -35.0\\
& 8 & 4368 & 6633 & -37.5\\
& 9 & 4364 & 6848 & -60.25\\
& 10 & 4109 & 6617 & -62.5\\
\bottomrule
\end{tabularx}
\noindent{\footnotesize{* Note that optimal solutions were -158.75, -76.75, -103.75, and -106.75.}}
\caption{Full QUBO results.\label{tab:qubo_results}}
\end{table}

\begin{table}[H]
\begin{tabularx}{\textwidth}{CCCCC}
\toprule
\textbf{Test No.} & \textbf{Repeat No.}	& \textbf{Circuit Depth} & \textbf{2-qubit gates} & \textbf{Found solution *}\\
\midrule
\multirow[m]{10}{*}{1}
& 1 & 2193 & 1079 & -152.0\\
& 2 & 2209 & 1099 & -152.0\\
& 3 & 2195 & 1078 & -152.0\\
& 4 & 2275 & 1102 & -152.0\\
& 5 & 2227 & 1085 & -152.0\\
& 6 & 2272 & 1086 & -152.0\\
& 7 & 2247 & 1081 & -152.0\\
& 8 & 2234 & 1107 & -152.0\\
& 9 & 2242 & 1102 & -152.0\\
& 10 & 2196 & 1087 & -152.0\\
\midrule
\multirow[m]{10}{*}{2}
& 1 & 5374 & 2888 & -74.5\\
& 2 & 5281 & 2868 & -74.5\\
& 3 & 5354 & 2931 & -74.5\\
& 4 & 5370 & 3006 & -74.5\\
& 5 & 5407 & 2913 & -73.0\\
& 6 & 5421 & 2906 & -74.5\\
& 7 & 5373 & 2911 & -74.5\\
& 8 & 5360 & 2884 & -74.5\\
& 9 & 5327 & 2926 & -74.5\\
& 10 & 5368 & 2882 & -74.5\\
\midrule
\multirow[m]{10}{*}{3}
& 1 & 13553 & 9225 & -101.5\\
& 2 & 13557 & 9146 & -100.0\\
& 3 & 13705 & 9171 & -100.0\\
& 4 & 13639 & 9199 & -99.25\\
& 5 & 13684 & 9199 & -101.5\\
& 6 & 13539 & 9189 & -100.0\\
& 7 & 13805 & 9274 & -100.0\\
& 8 & 13694 & 9189 & -101.5\\
& 9 & 13538 & 9200 & -100.0\\
& 10 & 13589 & 9159 & -99.25\\
\midrule
\multirow[m]{10}{*}{4}
& 1 & 23041 & 17204 & -102.25\\
& 2 & 23124 & 17108 & -102.25\\
& 3 & 23174 & 17115 & -102.25\\
& 4 & 23061 & 17049 & -102.25\\
& 5 & 23250 & 17151 & -102.25\\
& 6 & 23317 & 17109 & -102.25\\
& 7 & 23096 & 17068 & -102.25\\
& 8 & 23029 & 17058 & -102.25\\
& 9 & 23183 & 17127 & -102.25\\
& 10 & 23020 & 17136 & -103.0\\
\bottomrule
\end{tabularx}
\noindent{\footnotesize{* Note that optimal solutions were -152.0, -74.5, -101.5, and -104.5.}}
\caption{Full HUBO results (unfactored). \label{tab:hubo_results}}
\end{table}

\begin{table}[H]
\begin{tabularx}{\textwidth}{CCCCC}
\toprule
\textbf{Test No.} & \textbf{Repeat No.}	& \textbf{Circuit Depth} & \textbf{2-qubit gates} & \textbf{Found solution *}\\
\midrule
\multirow[m]{10}{*}{1}
& 1 & 1689 & 786 & -152.0\\
& 2 & 1712 & 803 & -152.0\\
& 3 & 1678 & 790 & -152.0\\
& 4 & 1723 & 787 & -152.0\\
& 5 & 1716 & 788 & -152.0\\
& 6 & 1686 & 791 & -152.0\\
& 7 & 1677 & 777 & -152.0\\
& 8 & 1714 & 782 & -152.0\\
& 9 & 1689 & 804 & -152.0\\
& 10 & 1735 & 787 & -152.0\\
\midrule
\multirow[m]{10}{*}{2}
& 1 & 3682 & 1841 & -74.5\\
& 2 & 3685 & 1868 & -74.5\\
& 3 & 3715 & 1856 & -74.5\\
& 4 & 3712 & 1845 & -74.5\\
& 5 & 3628 & 1882 & -74.5\\
& 6 & 3700 & 1852 & -74.5\\
& 7 & 3678 & 1855 & -74.5\\
& 8 & 3762 & 1854 & -74.5\\
& 9 & 3663 & 1873 & -74.5\\
& 10 & 3694 & 1860 & -74.5\\
\midrule
\multirow[m]{10}{*}{3}
& 1 & 9733 & 5568 & -101.5\\
& 2 & 9624 & 5449 & -100.0\\
& 3 & 9549 & 5464 & -100.0\\
& 4 & 9516 & 5546 & -100.0\\
& 5 & 9719 & 5510 & -101.5\\
& 6 & 9592 & 5534 & -100.0\\
& 7 & 9598 & 5492 & -101.5\\
& 8 & 9561 & 5487 & -99.25\\
& 9 & 9588 & 5474 & -99.5\\
& 10 & 9679 & 5528 & -100.0\\
\midrule
\multirow[m]{10}{*}{4}
& 1 & 16465 & 9778 & -102.25\\
& 2 & 16511 & 9788 & -104.5\\
& 3 & 16441 & 9772 & -102.25\\
& 4 & 16424 & 9788 & -102.25\\
& 5 & 16404 & 9791 & -102.25\\
& 6 & 16277 & 9796 & -102.25\\
& 7 & 16348 & 9753 & -103.0\\
& 8 & 16365 & 9736 & -102.25\\
& 9 & 16536 & 9793 & -102.25\\
& 10 & 16388 & 9764 & -102.25\\
\bottomrule
\end{tabularx}
\noindent{\footnotesize{* Note that optimal solutions were -152.0, -74.5, -101.5, and -104.5.}}
\caption{Full factored HUBO results. \label{tab:hubo_f_results}}
\end{table}

\end{document}